\documentstyle[manuscript,aps]{revtex}
\input epsf

\newcommand{\beq}{\begin{equation}}
\newcommand{\eeq}[1]{\label{#1} \end{equation}}
\newcommand{\beqar}{\begin{eqnarray}}
\newcommand{\eeqar}[1]{\label{#1} \end{eqnarray}}

\def\beq{\begin{equation}}
\def\eeq{\end{equation}}
\def\bea{\begin{eqnarray}}
\def\eea{\end{eqnarray}}
\def\R{{\bf R}}
\def\Z{{\bf Z}}
\def\RP{{\bf RP}}
\def\S{{\bf S}}

\def\c{\cite}
\def\r{\ref}

\def\ep{\varepsilon}

\def\S{{\bf S}}

\def\bce{\begin{center}}
\def\ece{\end{center}}

\def\De{\Delta}
\def\ep{\epsilon}

\def\La{\Lambda}

\def\R{{\bf R}}
\def\Z{{\bf Z}}
\def\RP{{\bf RP}}
\def\S{{\bf S}}

\begin{document}
\draft
\preprint{SNUTP-98-109, SOGANG-HEP 247/98, APCTP-98-023, hep-th/9811010}
\setcounter{page}{0}
\title{\Large\bf ${\cal N}=8$ SCFT and M Theory on 
$AdS_4 \times {\bf RP}^7$} 
\author{\large \rm 
Changhyun Ahn}
\address{ \it  Dept. of Physics, 
Kyungpook National University, 
Taegu 702-701, Korea \\
\tt ahn@kyungpook.ac.kr} 
\author{\large \rm
Hoil Kim} 
\address{\it Topology and Geometry Research Center, 
Kyungpook National University, 
Taegu 702-701, Korea \\
\tt  hikim@gauss.kyungpook.ac.kr }
\author{\large Bum-Hoon Lee
and  Hyun Seok Yang 
}
\address{ \it Department of Physics, 
Sogang University, 
Seoul 121-742, Korea \\
\tt bhl, hsyang@physics4.sogang.ac.kr}
\date{\today}

\maketitle

\begin{abstract}

We study M theory on $AdS_4 \times \RP^7$ corresponding to 3 dimensional 
${\cal N}=8$ superconformal field theory which is the strong coupling
limit of  
3 dimensional super Yang-Mills theory. 
 For $SU(N)$ theory, a wrapped M5 brane 
on $\RP^5$  can be 
interpreted as baryon vertex. 
For $SO(N)/Sp(2N)$ theory, by using the property of 
(co-)homology of $\RP^7$, 
we classify various wrapping branes and 
consider domain walls and the baryon 
vertex.
\end{abstract}

\setcounter{footnote}{0}
\newpage
\section{Introduction}
\setcounter{equation}{0}

In \cite{mal}  the large $N$ limit of superconformal field theories (SCFT) 
was described by taking the
supergravity limit on anti-de Sitter (AdS) space.
The scaling dimensions of operators of SCFT can be obtained from the 
masses of particles in string/M theory \cite{polyakov}. 
In particular, 
${\cal N}=4$ $SU(N)$ super Yang-Mills theory in 4 dimensions is described by
Type IIB string theory on $AdS_5 \times {\bf S}^5$.  
This AdS/CFT correspondence was tested by studying
the Kaluza-Klein (KK) states of supergravity theory and 
by comparing them with the chiral primary operators
of the SCFT on the boundary \cite{ot}. 
There exist also ${\cal N} =2, 1, 0$ superconformal theories in 
4 dimensions which have corresponding supergravity description 
on orbifolds of $AdS_5 \times {\bf S}^5$ \cite{kachru} and
the KK spectrum description on the twisted states of $AdS_5$ orbifolds
was discussed in \cite{gukov}.
The energy of quark-antiquark pair \cite{ryma}, 
glueball mass spectrum \cite{coot} 
and the energy of baryon as a
function of its size \cite{bisy,ima}
were analytically calculated based on this correspondence. 
The field theory/M theory duality also provides
a supergravity description on $AdS_4$ or $AdS_7$
for some superconformal theories
in 3 or 6 dimensions, respectively \cite{mal}. 
The maximally supersymmetric theories
have been studied in \cite{aoy,lr} and
the lower supersymmetric case was also realized on the worldvolume of 
M theory at orbifold singularities \cite{fkpz}. 

The gauge group of the boundary theory becomes 
$SO(N)/Sp(2N)$ \cite{witten1} by taking
appropriate orientifold operations for 
the string theory on $AdS_5 \times \S^5$ (See also \cite{aoy,kaku}). 
By analyzing the discrete torsion for $B$ fields, the possible models
of gauge theory are topologically classified and 
many features of gauge theory are described by various wrapping branes.
By generalizing the work of \cite{witten1} to the
case of $AdS_7 \times \RP^4$ where the eleventh dimensional
circle is one of $AdS_7$ coordinates, $(0,2)$ six dimensional SCFT
on a circle rather than uncompactified full M theory was described in
\cite{aky}.
For $SU(N)$ $(0,2)$ theory, a wrapped D4 brane on $\S^4$ together
with fundamental strings connecting a D4 brane on the boundary
of $AdS_7$ with the D4 brane on $\S^4$ was
interpreted as baryon vertex. By putting $N$ M5 branes in the
$\R^5/\Z_2$ orbifold singularity \c{hori}, where the ${\Z}_2$ acts by a
reflection of the 5 directions transverse to the M5 branes and also
by changing the sign of the 3-form field $C_3$, the large
$N$ limit of the $SO(2N)$ (0,2) SCFT and $\RP^4$ orientifold after
removing the $\R^5/\Z_2$ orbifold singularity were obtained.
Then, using the property of (co-)homology of $\RP^i \subset \RP^4$, 
various wrapping branes and their topological
restrictions were discussed.                                         

Recently, Sethi \cite{sethi} 
found that ${\cal N}=8$ $Sp(2N)$ and $SO(2N+1)$ gauge 
theories in 3 dimensions flow to the same strong coupling fixed point.
This was confirmed by turning on discrete torsion of M2 branes 
on $\R^8/\Z_2$. By evaluating $C_3 \wedge G_4$ over $\RP^7$ 
where $G_4 = dC_3$ is four-form field in M theory, 
it was shown that there exists M2 brane charge shift 
for the two types of orientifold two-plane.
This result implies that in IR limit three dimensional ${\cal N}=8$
SYM theories can flow to two distinct strong coupling conformal 
field theories. 

In this paper, we generalize the work of \cite{witten1,aky} to the
case of $AdS_4 \times \RP^7$ where the eleventh dimensional circle is one of
$\RP^7$ coordinates,\footnote{Although the eleven dimensional solution by 
uplifting the D2 brane solution is not exactly M2 brane solution in general, 
when the eleventh dimension is compact, by taking M2 branes to be 
localized in the eight transverse dimensions, 
they will resemble each other
more and more in M theory limit \cite{imsy,hks}.}
as we briefly mentioned this possibility 
in the previous paper \cite{aky}. 
In section II, it will be shown that for $SU(N)$ theory, 
a wrapped M5
brane on $\RP^5$ can be interpreted as baryon vertex \cite{fabbri}. 
By putting $N$ M2 branes in the 
$\R^8/\Z_2$ orbifold singularity, 
where the ${\Z}_2$ acts by a
reflection of the 8 directions transverse to the M2 branes, 
we will obtain the large
$N$ limit of ${\cal N}=8$ SCFT on $\RP^7$ orbifold \c{aoy}. 
Then, using the property of (co-)homology of 
$\RP^i \subset \RP^7$,
we classify various wrapping branes and discuss 
their topological restrictions. 
In section III, we consider domain walls and the baryon 
vertex. 
Finally, in section IV, we will discuss important open problems 
and comment on the future directions.

\section{ $SO/Sp$ SYM and Branes on ${\bf RP^7}$}
\setcounter{equation}{0}

\subsection{The Baryon Vertex in $SU(N)$}

Let us consider M theory on $\R^{3} \times \R^7 \times \S^1 $. 
For $\R^3$, we can take $(x^0, x^1, x^2)$ directions and for 
$\R^7$, we take $(x^3, x^4, x^5, x^6, x^7, x^8, x^9)$ 
transverse to M2 branes. The eleventh coordinate
$x^{10}$ is compactified on a circle $\S^1$ and is 
a periodic coordinate of period $2 \pi$. 
For small radius of $\S^1$, we can regard M theory as Type IIA
string theory which can be 
described in the context of $AdS_4 \times \S^7$ where
D2 branes are realized by transverse M2 branes. The radial function
$\rho= \sqrt{ \sum_{i=3}^{10} (x^i)^2} $ of $\R^7 \times \S^1$
will be one of the $AdS_4$ coordinates, 
the other three being the ones in $\R^3$. 

The $AdS_4 \times \S^7$ compactification has $N$ units of
seven-form flux on $\S^7$ as follows \cite{bisy}:
\bea
\int_{\S^7} \frac{G_7}{2 \pi}=N,
\label{n}
\eea 
where $G_7$ is seven-form field which is a hodge dual of
$G_4=dC_3$ four-form field. 
By putting a large number of $N$ of coincident M2 branes 
and taking the near horizon limit,
the metric becomes  that \cite{fabbri} of $AdS_4 \times \S^7$
\bea
ds_{11}^2 = \frac{r^4}{L^{2/3}} \eta_{\mu \nu} d y^{\mu} d y^{\nu} +
L^{1/3} \left(  \frac{dr^2}{r^2} +   g_{ij} dx^i dx^j \right).
\eea
The scale $L$ is related to $N$ by \cite{fabbri}
\bea
L= \left(\frac{\La}{6} \right)^{-3} =
\ell_p^6 2^5 \pi^2 N 
\label{scale}
\eea 
where $\ell_p$ is a Planck scale which is the only universal parameter
in M theory and $\mbox{Vol}(\S^7)= \frac{\pi^4}{3} (\frac{6}{\La})^{7/2}$. 
\footnote{The normalization \cite{fabbri} 
for four-form field strength is 
$G_{ijkl} = e \ep_{ijkl}$ where the parameter $e$ is a real constant. By 
plugging this into the 11 dimensional field equations, it leads to the 
product of 4 dimensional Einstein space, $R_{\mu \nu}= -
2 \La \eta_{\mu \nu}$ with Minkowski signature$(-,+,+,+)$ and 7 dimensional 
Einstein space $R_{ij}= \La g_{ij}$ where $\La$ is defined by $\La=
24 e^2/\kappa^{4/9}$ through grvitational constant $\kappa$. Moreover 
$\kappa^2= 8 \pi G_{11} = (2\pi)^8 \ell_p^9/2$.} 
The first equation arises when we write $AdS_4$ radius 
in terms of both cosmological constant $\La$ and scale factor $L$.
Since M2 branes
have the operators with dimension $\sqrt{N}$ by M2 tension formula
and M5 branes have the 
operators with dimension $N$ through the relation between mass, tension
\cite{dealwis}
and volume of branes,
we consider wrapping a M5 brane over 5-cycle of $\S^7$. 

A 5-cycle 
of minimum volume is to take the subspace at a constant value of 
two coordinates. For the 5 volume,
$\mbox{Vol(5-cycle)}$,
one can find 
$
\mbox{Vol(5-cycle)} = \mbox{Vol}(\S^5) =
\pi^3 \left( \frac{6}{\La} \right)^{5/2}.
$
The mass of the M5 brane wrapped over 5-cycle, given by M5 brane tension times
$\mbox{Vol(5-cycle)}$, is
\bea
m= \frac{1}{(2 \pi)^5 \ell_p^6} \mbox{Vol(5-cycle)}.
\label{mass}
\eea
By the relation     
$
m^2= \frac{2\La}{3} (\De-1)(\De-2) \approx  \frac{2\La}{3} \De^2
$
for large $\De$ and
the relations (\ref{mass}) and (\ref{scale}),
one  gets for the mass formula \cite{fabbri} for the dimension of a baryon
corresponding to the M5 brane wrapped 5-cycle
\bea
\De= \frac{\pi N}{\La} \frac{\mbox{Vol}(\S^5)}{\mbox{Vol}(\S^7)}=
\frac{N}{2}.
\label{nover2}
\eea

\subsection{The ${\bf RP^7}$ Orientifold}

It was observed \cite{aoy}
that the large $N$ limit of ${\cal N}=8$ SCFT in 3 dimensions
corresponds to $N$ M2 branes coinciding
at $\R^8/\Z_2$ orbifold singularity.
Let us consider M theory on
$\R^3  \times ( \R^7 \times \S^1 )/\Z_2$.\footnote{The eleven
dimensional spacetime is not a principal 
$U(1)$-bundle over 10 dimensional spacetime although it can be defined 
as a $U(1)$-bundle over 10 dimensional spacetime. 
However, it will still make sense to
consider RR $U(1)$ gauge field as one-form
with values in the twisted bundle. We thanks to K. Hori for pointing
out this.} 
The $\Z_2$ acts by sign change on 
all eight coordinates in $\R^7$ and $\S^1$ as follows:
$
(x^3,\cdots, x^{10}) \rightarrow 
(-x^3,\cdots, -x^{10}).
$ 
This gives two orbifold singularities at $x^{10}=0$ and $x^{10}=\pi$, 
each of which locally looks like $\R^8/\Z_2$.
The angular directions in $\R^8 /\Z_2$ are identified with
$\RP^7 $.
Consider $N$ parallel M2 branes which are sitting 
at an orbifold two-plane (O2-plane) 
which is located at $x^3=\cdots=x^{10}=0$. 
Note that there is another singularity at $x^{10}=\pi$ but we will
focus on the theory at the origin which has corresponding interacting 
superconformal field theory \cite{seiberg}. 
We will
describe how ${\cal N}=8$ SCFT in 3 dimensions
can be interpreted as M theory on $AdS_4 \times \RP^7$ where the eleventh 
dimensional circle is in $\RP^7$ space.  
This is our main goal in this paper.

Let us study the property of  $AdS_4 \times \RP^7$ orientifold.
Let $x$ be the generator of $H^1(\RP^7, \Z_2)$ which is isomorphic
to $\Z_2$, 
$\Sigma$ be a string
worldsheet and $w_1(\Sigma) \in H^1(\Sigma, \Z_2)$ be the obstruction
to its orientability. Then we only consider the map $\Phi: \Sigma \rightarrow
AdS_4 \times \RP^7$ such that $\Phi^*(x)=w_1(\Sigma)$.
Since $\Z_2$ action on $\S^7$ is free (no orientifold fixed points),
there is no open string sector. In the orientifold the string
world sheet need not be orientable and a basic case of 
an unorientable closed string worldsheet is $\Sigma=\RP^2$, 
which can be identified with the quotient of the two sphere $\S^2$
by the overall sign change.
The map $\Phi: \RP^2 \rightarrow \RP^7$ satisfying the constraints
$\Phi^*(x)=w_1(\RP^2)$ is the embedding 
$(x_1, x_2, x_3) \rightarrow (x_1, x_2, x_3, 0, 0, 0, 0, 0)
$.

In M theory, there is the Chern-Simons interaction in eleven dimensional 
supergravity,
\bea
\label{11cs}
- \frac{1}{24\pi^2} \int C_3 \wedge G_4 \wedge G_4. 
\eea
where $G_4=dC_3$.
A compactification of M-theory on an eightfold $X_8$ receives tadpole 
contribution for the $C_3$ three-form field in one-loop \cite{bb,svw} 
\bea
- \int_{X_8} C_3 \wedge I_8(R), 
\eea
where $J=-\int_{X_8}I_8(R)=\chi/24$, with $\chi$ the Euler characteristic 
of the eightfold $X_8$ and $I_8(R)$ is an eight-form constructed 
as a quartic polynomial in the curvature.
The condition for a consistent M theory 
compactification on an eightfold is thus
\bea
\label{eomc}
\frac{\chi}{24}-\frac{1}{8\pi^2}\int_{X_8} G_4 \wedge G_4 -n=0,
\eea
where $n$ is the number of M2 branes filling the vacuum. 
It can be deduced from the relation (\ref{eomc}) the orbifold $\R^8/\Z_2$ 
carries $-1/8$ units of M2 brane charge \c{djm}.\footnote{We count
the number of
brane charges on $\Z_2$ orbifold in the double cover of the 
$\Z_2$ quotient where the charge of a D2 brane is 1.}

It was shown by Sethi \cite{sethi} 
that there exist three O2-planes \footnote{We will also not 
consider nontrivial holonomy (RR $U(1)$ Wilson line) 
around the eleventh circle as in \cite{sethi} 
since the gauge theory has a smooth strong coupling limit 
in decompactified M theory limit and the holonomy indeed vanishes 
due to $H^2({\bf RP}^7, {\widetilde {\bf Z}})=0$.} 
\bea
i) & \; SO(2N)\; \mbox{with} \; O2^-: \;
& \mbox{M theory on ${\bf R}^3 \times ({\bf R}^7 \times 
{\bf S^1})/{\bf Z}_2$}, \nonumber\\
ii) & \; SO(2N+1)\; \mbox{ with } \; \widetilde{O2}^{+}: \; 
& \mbox{M theory on ${\bf R}^3 \times 
({\bf R}^7 \times {\bf S}^1)/{\bf Z}_2$},\nonumber\\
iii) & \; Sp(2N)\; \mbox{ with}\; O2^+: \; 
& \mbox{M theory on ${\bf R}^3 \times 
({\bf R}^7 \times {\bf S}^1)/{\bf Z}_2 $}.  
\eea   
Here $O2^{-}$ characterized by $\R^7/\Z_2$
has $-1/4$ unit of D2 brane charge. 
By promoting to M theory, each fixed point (0 and $\pi$ on the circle) 
realized by
$\R^8/\Z_2$ carries $-1/8$ unit of M2 brane charge.
We should have a single D2 brane stuck on the $O2^-$ plane to get $SO(2N+1)$
gauge group and the orientifold $\R^7/\Z_2$ carries 3/4 units 
of D2 brane charge.
It is denoted by ${\widetilde{O2}}^{+}$. In strong coupling limit, 
${\widetilde{O2}}^{+}$ plane splits into two orbifolds $\R^8/\Z_2$. 
Apparently the stuck M2 brane also splits into two fluxes, each
carrying $1/2$ unit of M2 brane charge. 
Each fixed point $\R^8/\Z_2$ has thus $3/8=
-1/8+1/2$ units of M2 brane charge.
As shown by Sethi \cite{sethi}, the M
theory realization of the charge shift is given by the term:
\bea
\label{cshift}
-\frac{1}{2}\int_{\RP^7}\frac{C_3}{2\pi}\wedge\frac{G_4}{2\pi}=
-\frac{1}{2}\int_{\cal M}\frac{G_4}{2\pi}\wedge\frac{G_4}{2\pi}=
\frac{1}{4},
\eea
where $G_4/2\pi$ is the torsion class and 
$\RP^7$ is the boundary of the smooth eightfold ${\cal M}$. 
Finally we can obtain $O2^{+}$ plane by the basis choice of the
Chan-Paton factors giving gauge group $Sp(2N)$,
which has the 1/4 unit of D2 brane charge.
In this case, the M theory interpretation on the orbifold singularity is 
a little different:
the singularity at $x^{10}= 0$ has the charge $3/8= -1/8+ 1/2$ units
of M2 brane where the $1/2$ (in the double cover $\S^7$) charge shift is
realized by turning on the discrete torsion as in Eq. (\r{cshift})
while the singularity
at $\pi$ having $-1/8$ units of M2 brane charge. Thus we wish to classify the
O2-planes in terms of distinct fluxes for the four-form $G_4$ on
$\RP^7$ at the origin which correspond to distinct strong coupling
limits for O2 planes \c{sethi}. The relevant cohomology corresponds to the
possible choices of discrete torsion and is given by
\beq
\label{ch4}
H^4(\RP^7,\Z)=\Z_2.
\eeq
Consequently, $O2^-$ plane flows to the case without discrete torsion 
while $\widetilde{O2}^{+}, O2^{+}$ flow to the case with discrete torsion.

Since the $SO(N)/Sp(2N)$ gauge theories for large $N$ 
can be distinguished by the
sign of the string ${\bf RP}^2$ diagram,
this can be classified by the discrete torsion 
of the $B_{NS}$ field \c{witten1}.
Note that the orbifolding in $\R^8/\Z_2$ does not act
on $C_3$ but it acts on $x_{10}$. 
Since the $B_{\mu\nu}$ field of Type IIA string theory corresponds in
M theory to $C_{\mu\nu 10}$,
${\bf Z}_2$ action in ten dimensions flips the sign of $B_{NS}$. 
This means that a cohomology class $[H_{NS}]$ takes values in a
twisted integer coefficient ${\widetilde {\bf Z}}$ where the twisting is
determined by an orientation bundle. 
The relevant cohomology groups measuring the topological types of the fields 
$B_{NS}$ is given by 
\bea
\label{ch3}
H^3({\bf RP}^7, {\widetilde {\bf Z}})\approx {\bf Z}_2.
\eea 
Thus the relevant cohomology groups
measuring the topological types for the O2 planes are given by
\bea
H^3({\bf RP}^7, {\widetilde {\bf Z}})\approx {\bf Z}_2,\qquad
H^4({\bf RP}^7, {\bf Z})\approx {\bf Z}_2.
\eea 
If we denote the values of the cohomologies 
$H^3({\bf RP}^7, {\widetilde {\bf Z}})$ and $H^4({\bf RP}^7, {\bf Z})$ as
$(\alpha, \beta)$ respectively, 
we get the topological classification of the three models:
\begin{eqnarray} 
O2^- :(\alpha, \beta)=(0,0),\qquad  
\widetilde{O2}^+ :(\alpha, \beta)=(0,1),\qquad 
O2^+ :(\alpha, \beta)=(1,1).
\end{eqnarray}
Note that the topological type of $O2^+$ at $x^{10}=\pi$ is 
$(\alpha, \beta)=(1,0)$.

\subsection{Various Wrapped Branes}

Now we consider the possibilities of brane wrapping on $\RP^7$ in the
M theory. 
The wrappings of M2 brane 
unwrapped around $x^{10}$ and M5 brane wrapped around $x^{10}$ 
are classified 
by the ordinary (untwisted) homology 
$H_i({\bf RP}^7, {\bf Z})$ for wrapped on an
$i$-cycle in ${\bf RP}^7$ since the twobrane charge is even under the
orientifolding operation (it comes from the fact that, in M theory, 
${\bf Z}_2$ action does not act on the three-form $C_3$) and the
fivebrane is dual to the twobrane:\\
$(i)$ unwrapped M2 brane, giving a twobrane in $AdS_4$,\\
$(ii)$ wrapped on a one-cycle, to give a onebrane 
in $AdS_4$, classified by $H_1({\bf RP}^7,{\bf Z})={\bf Z}_2$,\\ 
The unwrapped M5 brane is not possible. \\
$(iii)$ wrapped on a one-cycle, to give a fourbrane 
in $AdS_4$, classified by $H_1({\bf RP}^7,{\bf Z})={\bf Z}_2$,\\
$(iv)$ wrapped on a three-cycle, to give a twobrane 
in $AdS_4$, classified by $H_3({\bf RP}^7, {\bf Z})={\bf Z}_2$.\\ 
$(v)$ wrapped on a five-cycle, to give a zerobrane 
in $AdS_4$, classified by $H_5({\bf RP}^7, {\bf Z})={\bf Z}_2$.\\ 
The wrappings of M2 brane wrapped around $x^{10}$ and
M5 brane unwrapped around $x^{10}$ 
are classified 
by the twisted homology 
$H_i({\bf RP}^7, {\bf \widetilde{Z}})$ for wrapped on an
$i$-cycle in ${\bf RP}^7$.
The wrapping modes that would give onebranes are not possible
since $H_1({\bf RP}^7, {\widetilde {\bf Z}})=0$.\\
The unwrapped M5 brane (unwrapped around $x^{10}$) is not possible. \\ 
$(vi)$ wrapped on a two-cycle, to give a threebrane 
in $AdS_4$, classified by $H_2({\bf RP}^7, {\widetilde {\bf Z}})={\bf Z}_2$,\\
$(vii)$ wrapped on a four-cycle, to give a onebrane 
in $AdS_4$, classified by $H_4({\bf RP}^7, {\widetilde {\bf Z}})={\bf Z}_2$. \\

The $k$ units of KK momentum mode around the 
eleventh circle ${\bf S}^1$ can be identified as $k$ D0-branes which
is charged BPS particles.\footnote{Note that the D0 brane charge
is odd under the $\Z_2$ action since $A_{\mu}^{RR}=G_{\mu 10}$
and the $\Z_2$ flips the orientation of $\S^1$ and thus 
the RR $U(1)$ gauge field should be considered as twisted one-form.} 
In the three dimensional gauge theory context \cite{seiberg}, 
a new scalar modulus appears as the dual of a photon, 
the magnetic scalar photon, coming from dualizing the vector 
in three dimensions. This corresponds to the expectation value 
of the localized eleven dimensional coordinate
of a M2 brane. Then the effect of D0 branes exchange 
between M2 branes can be captured by instanton effects
in 3-dimensional SYM and renders it $SO(8)$ invariant \cite{hks}. 

According to the similar arguments done in Type IIB description 
\cite{witten1} and Type IIA description \c{aky}, 
one can derive a 
topological restriction on the brane wrappings on ${\bf RP}^7$ 
just described. In particular, since 
the topological restriction coming from
the holonomy of the connection $A_{RR}$ on the $U(1)$-bundle
would be not considered for the reason explained in the previous
footnote, it is sufficient only to consider the discrete torsion 
$\theta_{NS}$ of the field $B_{NS}$. We will show that 
the description of brane wrappings on ${\bf RP}^7$ 
is consistent with the topological
restriction coming from the RR discrete torsion $\theta_{RR}$, which
should vanish in our case.
In the case $(ii)$ and $(iii)$, there is no restriction
on wrapping of M2 and M5 branes on ${\bf RP}^1 \subset {\bf RP}^7$,
since in this case ${\bf RP}^2$ cannot even be deformed into the M2 or
M5 brane. 
In the case $(iv)$, the M5 brane can be wrapped on 
${\bf RP}^3$, to make a twobrane in $AdS_4$, 
only if $\theta_{NS} \neq 0$, 
since $H^2({\bf RP}^3, {\bf Z})=\Z_2$.
In the case $(v)$, the M5 brane can be wrapped on 
${\bf RP}^5$, to make a zerobrane in $AdS_4$, 
only if $\theta_{NS} \neq 0$, 
since $H^2({\bf RP}^5, {\bf Z})=\Z_2$.
In the case $(vi)$, there is no restriction on wrapping of
M5 branes on ${\bf RP}^2 \subset {\bf RP}^7$, to make a threebrane in
$AdS_4$, since $H^2({\bf RP}^2, {\widetilde {\bf Z}})={\bf Z}$.
In the case  $(vii)$, the M5 brane can be wrapped on 
${\bf RP}^4$, only if $\theta_{RR}=0$, 
since $H^2({\bf RP}^4, {\widetilde {\bf Z}})=0$.

\section{ Gauge Theory and Branes on $\RP^7$}
\setcounter{equation}{0}

\subsection{Domain Walls}

Let us consider the objects in $AdS_4\times {\bf S}^7$ and 
$AdS_4\times {\bf RP}^7$ that look like twobranes in the four 
noncompact dimensions of $AdS_4$. Since the $AdS_4$ has three spatial
dimensions, the twobrane could potentially behave as a domain wall,
with the ``jumping'' as one crosses the
twobrane. In $AdS_4\times {\bf S}^7$ and $AdS_4\times {\bf RP}^7$, 
the only such objects are M2 brane in the case $(i)$ 
and the M5 brane in the case $(iv)$ in section IIC. 
Note that, in the case of $AdS_4\times {\bf S}^7$, the M5 brane can not
wrap on a five-cycle in $\S^7$ since $H_5(\S^7, \Z)=0$, so does not
give rise to a domain wall in $SU(N)$ gauge theory. 

Since the twobrane is the electric source of the four-form field $G_4$, 
the integrated four-form flux over ${\bf S}^7$ 
or ${\bf RP}^7$ jumps by one unit when one
crosses the twobrane. This means that the gauge group of the boundary
conformal field theory can change, for example, from $SU(N)$ on one side to 
$SU(N\pm 1)$ on the other side for $AdS_4\times {\bf S}^7$. 
For $AdS_4\times {\bf RP}^7$, 
if one is crossing the M2 brane, it changes from $SO(N)$ to $SO(N\pm 2)$ 
or from $Sp(N/2)$ to $Sp(N/2\pm 1)$ since the M2 brane charge changes 
by two units on double cover. 

The similar situation also occurs in the case of the M5 brane wrapped
on $\RP^3 \subset \RP^7$ to make a twobrane. Let $P$ and $Q$  be
points on opposite sides of the twobrane. Let $X$ be the
three-manifold $X=T\times \RP^4$, with $T$ a path from $P$ and $Q$,
intersecting the twobrane once. Since a generic $\RP^3$ and $\RP^4$ in
$\RP^7$ have one point of intersection, $X$ generically intersects the
M5 brane at one point. The boundary of $X$ is the union of the
two-manifolds $P\times \RP^4$ and $Q \times \RP^4$. 
Although it causes no change
of the gauge group in the boundary theory, this may have an important
effect on instanton corrections in field theory.

\subsection{The Baryon Vertex in $SO(N)/Sp(2N)$}

The baryon vertex in $SU(N)$ was obtained by wrapping a M5 brane over 
${\bf S}^5$. By analogy, one expects that the baryon vertex in 
$SO(N)$ or $Sp(2N)$ will consist of a M5 brane wrapped on ${\bf RP}^5$. 
If we are considering $SO(2k)$ gauge theory, there are $k$ units of 
five-form flux on ${\bf RP}^5$ 
when the M5 brane wraps once on ${\bf RP}^5$.
But there is no gauge invariant combination, in $SO(2k)$ gauge theory,
of $k$ external quarks to obtain a ``baryon vertex''. 
The baryon vertex of $SO(2k)$ gauge theory should couple 
$2k$ external quarks, not $k$ of them. 

Let $\Phi$ be the map of M5 brane worldvolume $X$ to 
$AdS_4\times{\bf RP}^7$. We must impose the condition that the
$B_{NS}$ fields should be topologically trivial when pulled back to
$X$ as implied in deriving the topological restrictions on the brane
wrapping in section II.  
If we choose the M5 brane topology as ${\bf S}^5$, 
the topological triviality of the field $B_{NS}$ is automatically obeyed 
since $H^3({\bf S}^5, \Z)=0$, the AdS baryon vertex thus
exists regardless of the gauge group of the boundary theory. 
Then the map $\Phi:{\bf S}^5 \rightarrow {\bf RP}^5$ gives 
the degree two map, in other words,   
${\bf S}^5$ wraps twice around ${\bf RP}^5$. Thus we can obtain the
correct baryon vertex coupling $2k$ quarks. 

In $Sp(k)$ gauge theory, a baryon vertex can decay to $k$ mesons \c{wittbr}. 
Thus, one may expect no topological stability for the $AdS_4$ baryon
vertex when $\theta_{NS} \neq 0$. 
However, since the stability of baryon 
in $SO(k)/Sp(k)$ gauge theory actually
can be encoded by the homology,
$H_5({\bf RP}^7, \Z)=\Z_2$, which implies the topological
stability of a baryonic charge, we get the field theory result. 
We also should consider the
possibility on an existence of nontrivial torsion class of 
the $B_{NS}$ fields due to the topology of the M5 brane 
worldvolume $X$, which is denoted as
$W\in H^3(X, \Z)$ \c{witten1}. 
Then this means that the correct global restriction is not that 
$i^*([H_{NS}])=0$ but rather that
\beq
i^*([H_{NS}])=W,
\eeq
where $i$ is the inclusion of $X$ in spacetime and $[H_{NS}]$ the
characteristic class of the $B_{NS}$ field. 
A possible $W$ can be determined by using 
the ``connecting homomorphism'' in an exact sequence of cohomology groups 
from the second Stieffel-Whitney class $w_2(X)\in H^2(X, {\Z}_2)$,
which means that the proper global restriction is $i^*([H_{NS}])=W$, 
i.e., $\theta_{NS} \neq 0$ which is also consisitent with the topological
anaylsis we did last section. 
If the baryon vertex decays via compact six-manifold $X$, we will find
that $W \neq 0$ since $w_2(X) \neq 0 $. 
Consequently, the brane decay via compact sixfold $X$ is possible only 
in $Sp(k)$ gauge theory. 
 
In $SO(N)$ gauge theory where $N$ is even or odd, 
super Yang-Mills theory with 16 supercharges
actually has $O(N)$ symmetry, not just $SO(N)$. The generator $\tau$
of the quotient $O(N)/SO(N) = {\bf Z}_2$ behaves as a global
symmetry. Since the baryon is odd under $\tau$, it cannot decay to
mesons which is even under $\tau$. 
If there is a Pfaffian-like state which is odd under $\tau$, 
the possible decay channel may be mesons plus a Pfaffian 
as in \c{witten1}.
However, in our case, there is no definite candidate being role of the
Pfaffian. It is well-known \c{wittbr} that
two baryons in $SO(N)$ gauge theory can
annihilate into $N$ mesons. This is also consistent with the fact that, 
in $O(N)$, a product of two epsilon symbols can be rewritten as a sum
of products of $N$ Kronecker deltas. 
Their annihilation can be realized by the similar (and more simple) 
process to the pair annihilation of two fat strings discussed in \c{witten1}. 
That is, two identical M5 branes, whose worldvolumes are of 
the form $C\times \S^5$ and $C^\prime \times \S^5$ where 
$C$ and $C^\prime$ are timelike path, collide at any time 
where $C$ and $C^\prime$ coincide. 
These results imply that the decay of baryon 
in $SO(N)/Sp(N)$ gauge theory should belong to the element of 
$H_5(Y, \Z)=\Z_2$. When we consider $Y=\RP^7$, the baryon
vertex is stable since $H_5(\RP^7, \Z)=\Z_2$. 

\section{Discussion }
\setcounter{equation}{0}

To summarize, for $SU(N)$  theory,
we interpreted the baryon vertex as a wrapped M5 brane in $\S^7$.
When we go $SO(N)/Sp(2N)$ theory, $\R^8/\Z_2$ orbifold singularity was 
crucial to understand M theory realization of three types of O2 plane.
We constructed the possible brane wrappings on $\RP^7$ 
and determined their topological restrictions in each case.  
According to this classification, it was possible to interpret 
various wrapping branes on $\RP^7$
in terms of domain walls and the baryon 
vertex where the topological properties on $\RP^7$ are used.  

It was pointed out by Witten \c{private}
that the infrared limit of SYM on $\R^3$
related to ${\cal N}=8$ SCFT is quite subtle and there may be
suprising features in the topology in the infrared, such as
conservation laws that hold in the infrared but not exactly.
It may become important in the limit to consider the nonperturbative
instanton corrections in SYM theory, which is essential to recover $SO(8)$
invariance or eleven dimensional Lorentz invariance \c{hks}.
  
It was observed by Sethi \c{sethi} that M-theory realization of
O2 planes gives an amusing interpretation on the shift of membrane
charge by the discrete torsion. This interesting phenamenon may be
more clearly understood in terms of the method applied to O4 planes by
Gimon \cite{gimon} where, applying the T-S-T transformations, 
the O4 planes are related to O3 palnes which is more well understood.
By applying the same strategy, one can relate the O2 planes to
O3 planes by the T-S-T transformations where T-duality is taken along
the one direction of the O3 plane. We hope this work will be
accomplished near future and provide a new understanding on O2 planes.

Klebanov and Witten \cite{kw} found $AdS_5 \times T^{1,1}$ model which
is an example of holographic theory on a compact manifold which is not
locally $\S^5$ and the corresponding quantum field theory
can not be obtained
from the projection of maximal ${\cal N}=4$ theory. See also recent 
papers \cite{morr}.
It is well known that
there exist various types of seven dimensional compact Einstein manifold 
$X_7$ which is not locally $\S^7$. 
It would be interesting to study whether one can find
wrapping branes over cycles of $X_7$ and discuss their field theory 
interpretation.

\vspace{2cm}

{\bf Note added}: After this work has been finished, we found the
reference \cite{bk} which treats related subject.

\centerline{\bf Acknowledgments} 

We thank O. Aharony, K. Hori and E. Witten for email correspondence.
CA thanks K. Oh and R. Tatar for discussions on relating subjects.
This work is supported (in part) by the Korea Science 
and Engineering Foundation (KOSEF) through the Center 
for Theoretical Physics (CTP) at Seoul National University. 
BHL and HSY are also partially supported by the Korean Ministry of Education 
(BSRI-98-2414) and HK is supported by TGRC-KOSEF.
We thank Asia Pacific Center for Theoretical Physics (APCTP) for
hospitality where this work has been done. 



\begin{thebibliography}{99}
\bibitem{mal} J. Maldacena, Adv. Theor. Math. Phys. {\bf 2} (1998) 231.
\bibitem{polyakov} S. S. Gubser, I. R. Klebanov, A. M. Polyakov, 
Phys. Lett. {\bf B428} (1998) 105; E. Witten, 
Adv. Theor. Math. Phys. {\bf 2} (1998) 253.
\bibitem{ot} Y. Oz and J. Terning, hep-th/9803167.
\bibitem{kachru} S. Kachru and E. Silverstein, Phys. Rev. Lett. 
{\bf 80} (1998) 4855; A. Lawrence, N. Nekrasov and C. Vafa, hep-th/9803015.
\bibitem{gukov} S. Gukov, hep-th/9806180.
\bibitem{ryma} S.-J. Rey and J. Yee, hep-th/9803001; 
J. Maldacena, Phys. Rev. Lett. {\bf 80} (1998) 4859;
D. J. Gross and H. Ooguri, Phys. Rev. {\bf D58} (1998) 106002.
\bibitem{coot} C. Cs\'aki, H. Ooguri, Y. Oz and J. Terning,
hep-th/9806021;
R. de Mello Koch, A. Jevicki, M. Mihailescu, and J. P. Nunes, 
Phys. Rev. {\bf D58} (1998) 105009.
\bibitem{bisy} A. Brandhuber, N. Itzhaki, J. Sonnenschein and S. 
Yankielowicz, J. High Energy Phys. {\bf 07} (1998) 020.
\bibitem{ima} Y. Imamura, hep-th/9806162.
\bibitem{aoy} O. Aharony, Y. Oz and Z. Yin, Phys. Lett. {\bf B430} (1998) 87.
\bibitem{lr} R. G. Leigh and M. Rozali, Phys. Lett. {\bf B431} (1998) 311;
S. Minwalla, J. High Energy Phys. {\bf 10} (1998) 002;
E. Halyo, J. High Energy Phys. {\bf 04} (1998) 011; J. Gomis, hep-th/9803119.
\bibitem{fkpz} S. Ferrara, A. Kehagias, H. Partouche and A. Zaffaroni, 
Phys. Lett. {\bf B431} (1998) 42; M. Berkooz, hep-th/9802195;
R. Entin and J. Gomis, Phys. Rev. {\bf D58} (1998) 105008;
C. Ahn, K. Oh and R. Tatar, hep-th/9806041;
C. Ahn, K. Oh and R. Tatar, hep-th/9804093, 
to appear in Phys. Lett. {\bf B}.
\bibitem{witten1} E. Witten, J. High Energy Phys. {\bf 07} (1998) 006.
\bibitem{kaku} Z. Kakushadze, Nucl. Phys. {\bf B529} (1998) 157;
Phys. Rev. {\bf D58} (1998) 106003;
A. Fayyazuddin and M. Spalinski, hep-th/9805096;
O. Aharony, A. Fayyazuddin and J. Maldacena,
J. High Energy Phys. {\bf 07} (1998) 013;
C. Ahn, K. Oh and R. Tatar, hep-th/9808143;
S. S. Gubser and I. R. Klebanov, hep-th/9808075.
\bibitem{aky} C. Ahn, H. Kim and H. S. Yang, hep-th/9808182.
\bibitem{hori} K. Hori, hep-th/9805141.
\bibitem{sethi} S. Sethi, hep-th/9809162.
\bibitem{fabbri} D. Fabbri, P.Fre, L. Gualtieri, C. Reiner, A. Tomasiello,
A. Zaffaroni and A. Zampa, hep-th/9907219.
\bibitem{imsy}N. Itzhaki, J. M. Maldacena, J. Sonnenschein and 
S. Yankielowicz, Phys. Rev. {\bf D58} (1998) 046004.
\bibitem{hks}J. Polchinski and P. Pouliot, Phys. Rev. {\bf D56} (1997)
6601; E. Keski-Vakkuri and P. Kraus, hep-th/9804067; 
S. Paban, S. Sethi and M. Stern, hep-th/9808119; 
S. Hyun, Y. Kiem and H. Shin, hep-th/9808183.
\bibitem{dealwis} S.P. de Alwis, Phys.Lett. {\bf B388} (1996) 291.
\bibitem{seiberg} N. Seiberg, hep-th/9705117.
\bibitem{bb}K. Becker and M. Becker, Nucl. Phys. {\bf B477} (1996) 155.
\bibitem{svw}S. Sethi, C. Vafa and E. Witten, Nucl. Phys. {\bf B480} 
(1996) 213; E. Witten, J. Geom. Phys. {\bf 22} (1997) 1.
\bibitem{djm}K. Dasgupta, D. P. Jatkar and S. Mukhi, 
Nucl. Phys. {\bf B523} (1998) 465.
\bibitem{wittbr}E. Witten, Nucl. Phys. {\bf B223} (1983) 433.
\bibitem{private} E. Witten, private communication.
\bibitem{gimon} E. G. Gimon, hep-th/9806226.
\bibitem{kw} I. R. Klebanov and E. Witten, hep-th/9807080.
\bibitem{morr} D. R. Morrison and M. R. Plesser, hep-th/9810201;
K. Oh and R. Tatar, hep-th/9810244; C. P. Boyer and K. Galicki, 
hep-th/9810250.
\bibitem{bk} M. Berkooz and A. Kapustin, hep-th/9810257.


\end{thebibliography}
\end{document}